\documentclass[aps,pra,amsmath, amsfonts, amssymb, superscriptaddress, twocolumn, showpacs]{revtex4}

\usepackage{graphicx}
\usepackage{setspace}
\usepackage{wasysym}
\usepackage{color}
\usepackage{nicefrac}

\def\a{s}
\def\b{s}
 
\newcommand{\add}[1]{\if\a\b{{\color{red} #1}}\else{#1}\fi}
\newcommand{\comm}[1]{\if\a\b{{\color{blue} #1}}\else{#1}\fi}

\newcommand{\citeasnoun}[1]{Ref.~\onlinecite{#1}}

\renewcommand{\eqref}[1]{Eq.~(\ref{#1})}

\newcommand{\eqreftwo}[2]{Eqs.~(\ref{#1}) and~(\ref{#2})}

\newcommand{\figref}[1]{Fig.~\ref{fig:#1}}

\renewcommand{\Re}{\operatorname{Re}}
\renewcommand{\Im}{\operatorname{Im}}

\renewcommand{\section}[1]{\emph{#1}:}

\begin{document}
\def\linefigwidth{0.5\textwidth}
\def\smalllinefigwidth{0.35\textwidth}
\def\smallerlinefigwidth{0.25\textwidth}
\def\largelinefigwidth{0.5\textwidth}

\title{Delay--bandwidth and delay--loss limitations for
  cloaking of large objects}



\author{Hila Hashemi}
\affiliation{Department of Mathematics, Massachusetts Institute of Technology, Cambridge, MA 02139}
\author{Baile Zhang}
\affiliation{Singapore--MIT Alliance for Research and Technology Centre, Singapore 117543} 
\author{J.~D.~Joannopoulos}
\affiliation{Department of Physics, Massachusetts Institute of Technology, Cambridge, MA 02139}
\author{Steven~G.~Johnson}
\affiliation{Department of Mathematics, Massachusetts Institute of Technology, Cambridge, MA 02139}

\begin{abstract}
We show that the difficulty of cloaking is fundamentally limited by
delay--loss and delay--bandwidth/size limitations that worsen as the
size of the object to be cloaked increases relative to the wavelength,
using a simple model of ground-plane cloaking. These limitations
must be considered when scaling experimental cloaking
demonstrations up from wavelength-scale objects.
\end{abstract}
\pacs{78.67.Pt, 42.81.Dp}

\maketitle 


We will argue that the problem of cloaking becomes intrinsically more
difficult as the size of the object to be cloaked increases compared
to the wavelength, and is ultimately limited by fundamental
considerations involving the delay--bandwidth and delay--loss
products, even for ground-plane cloaks~\cite{Zhang08, Li08, Liu09}
where bandwidth is not limited by causality constraints.  The
difficulty is greatest for cloaking objects many wavelengths in
diameter (unlike experiments cloaking wavelength-scale
objects~\cite{Liu09, Schurig06, Kante08, Ma09OE, Liu09APL,
  Gabrielli09, Valentine09, Smolyaninov08, ErginSt10}), but
unfortunately this is the most useful regime for resolving an object
of interest. We illustrate these limitations with an idealized
one-dimensional~(1d) system in which cloaking is much simpler than in
three dimensions~(3d)---only one incident wave need be
considered---but in which the same limitations appear. We argue
  that the results and conclusions from this simplified model
apply even more strongly to 2d and 3d, and are consistent with
recent numerical calculations for 3d cloaks~\cite{Baile09}. We
conclude that cloaking of human-scale objects is challenging at radio
frequencies (RF), while cloaking such objects at much shorter
(e.g. visible) wavelengths is rendered impractical by the delay--loss
product. Despite the simplicity of this analysis, we arrive at
fundamental criteria that may help guide future research on the
frontiers of cloaking phenomena.

There has been intensive interest in cloaking, both theoretically and
experimentally, since the inspiring original papers describing how
coordinate transformations, mapped into inhomogeneous materials
(``transformation optics'')~\cite{Ward96} could theoretically render
an object invisible~\cite{Pendry06, Leonhardt06}. Since then, many
authors have proposed variations on the original cloaking
designs~\cite{Schurig06OE, Cummer06, Chen07PRL, Huang07, Zolla07,
  Cai07NP, Chen07PRB,Yan07, Rahm08, Kwon08, Jiang08, Kante08, Ma09,
  popa09, Cai07APL, Qiu09, Zhang08, Li08,
  kallos09,Leonhardt09}, and there have also been attempts
at experimental realization~\cite{Schurig06,Kante08, Liu09APL,
  Smolyaninov08, Smolyaninov09PRL, Liu09,
  Ma09OE,Gabrielli09,Valentine09, ErginSt10}. Most theoretical work,
however, has considered only lossless materials. In experiments,
significant reductions in the scattering cross-section (partial
cloaking) have been demonstrated mainly for objects on the scale of
the wavelength, with one recent exception~\cite{Smolyaninov09PRL}
discussed below.  Two practical concerns about cloaking have been
bandwidth limitations and the impact of losses/imperfections, and we
argue that these two difficulties become fundamentally more
challenging as the size of the object to be cloaked increases.

\begin{figure}[t]
 \centering
 \includegraphics[width=0.2\textwidth]{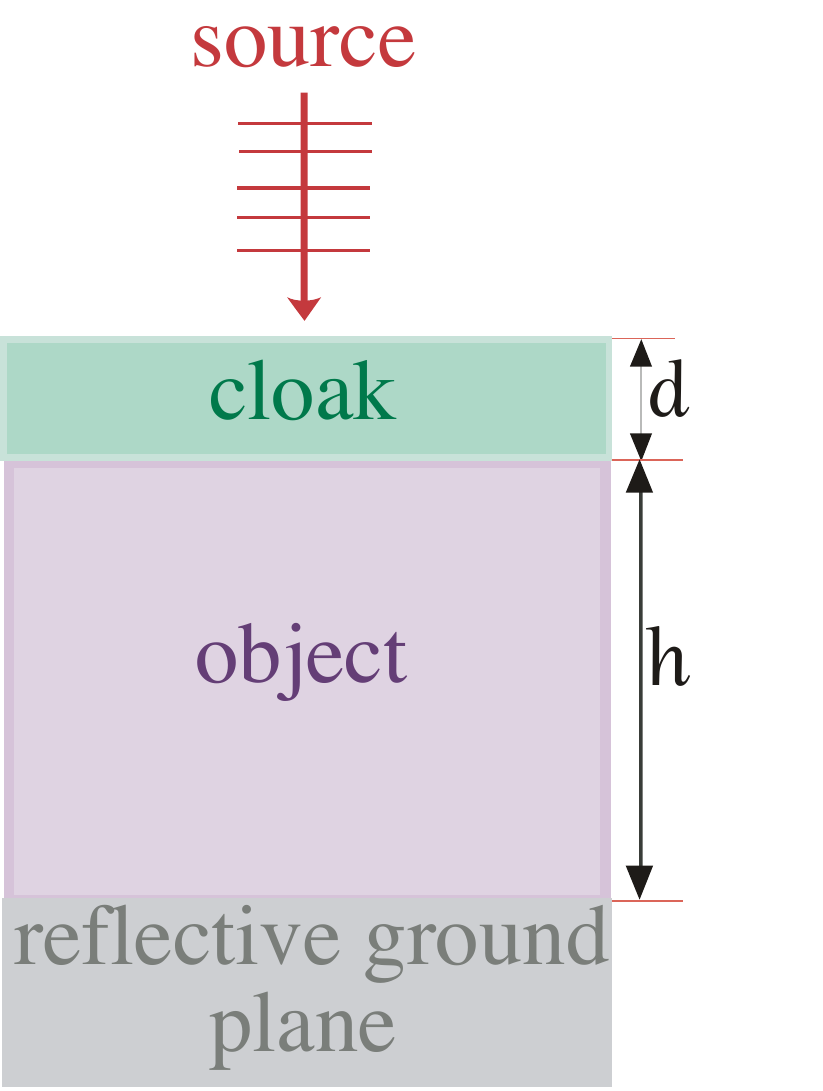}
 \caption{A $1$d ground-plane cloak}
 \label{fig:1d-cloak}
\end{figure}

Pendry pointed out that perfect cloaking in air/vacuum is impossible
over nonzero bandwidth, because rays traveling around the object must
have velocity $>c$ to mimic empty space~\cite{Pendry06}; this can be
interpreted as a causality constraint~\cite{Miller06}, and suggests a
causality limit on bandwidth even for imperfect cloaking.  However, it
was subsequently proposed that such bandwidth limitations are removed
for a ground-plane cloak, in which an object is hidden by a coordinate
transformation mapping it into a ground plane or
substrate~\cite{Zhang08, Li08, Liu09}. Causality constraints do not
seem to apply to ground-plane cloaks, because the reflected wave
travels a shorter distance in the presence of the cloak and hence does
not need a speed $> c$ to simulate absence of the object. Although
this design makes cloaking easier in both theory and practice, we
argue that even ground-plane cloaking is subject to
delay--bandwidth/size and delay--loss limitations that become more
stringent as the size of the cloaked object increases.  (To our
knowledge, experimental demonstrations of partial ground-plane
cloaking have thus far utilized only wavelength-scale
objects~\cite{Liu09, Ma09OE, Gabrielli09, Valentine09, ErginSt10}.)

\section{A simple 1d model}
In order to understand the limitations of ground-plane cloaking, we
consider the simplest possible circumstance: a 1d cloak to hide an
object of thickness $h$ on top of a substrate (e.g. a conducting
plane) in vacuum. This problem is conceptually much easier than
general cloaking, in that only a single incident (and reflected) wave
need be considered.  In contrast, even two-dimensional cloaking is far
more complex: not only would the object need to be cloaked from
incident waves at all angles, but for incident waves parallel to the
ground the cloaking problem becomes more similar to that of cloaking
an isolated object---with the associated causality constraints---as
the height of the object increases.  Since 1d cloaking appears to be so much
easier, any fundamental limitations that arise in this case should
apply even more strongly in~2d and~3d.

In this idealized 1d case, the cloak consists of some arbitrary
materials in a region of thickness $d$ on top of the object, as
depicted in \figref{1d-cloak}. We assume that the ground plane
reflects light with negligible loss in the bandwidth of interest (in
the trivial case of a black ground plane, one would merely need a
black cloak).  The function of the cloak is now simple: the cloak must
reflect incident waves with a delay equal to the time (and phase)
delay $\tau_0 \geq 2(h+d)/c$ that the reflected wave would incur in
the absence of the cloaked object. A similar delay must also be
achieved in 2d/3d cloaking for beams at any angle---the cloak
  must simulate the delay from bouncing ``through'' the object off the
  ground plane, and in fact the required delay increases for more
  oblique incidence (longer paths through the object). To be more
precise, suppose that the reflected wave from the bare ground plane,
at a height $h+d$, has a phase $\phi(\omega) \approx
\phi(\omega_0) + \phi'(\omega_0) (\omega - \omega_0)$ near some
frequency $\omega_0$, where the derivative $\phi'(\omega_0) =
\tau_0$ is the time delay~\cite{Lenz01}.  There are two cases.
First, the \emph{phase-delay case}: if the bandwidth is narrow,
so that $\phi'(\omega_0) (\omega - \omega_0)$ can be neglected, then
the cloak merely need achieve the correct phase
$\phi(\omega_0)$, but this imposes a bandwidth constraint: the
delay--bandwidth product $\tau_0\Delta\omega$ must be small.
(This corresponds to incident pulses of duration
$\sim 1/\Delta\omega \gg \tau_0$.)  Second, the \emph{time-delay
  case}: if the delay--bandwidth product is not small, then
$\phi'(\omega_0) (\omega - \omega_0)$ cannot be neglected and the
cloak must achieve a true time delay $\tau_0$ (an $\omega$-dependent
phase).  This raises two additional
difficulties.  First, it is well known that the achievable
delay--bandwidth product in finite-size passive linear systems is
limited~\cite{Lenz01, Tucker05, Miller07}.  Second, a long dwell time
in the cloak means that loss in the cloak must be
small.  We deal with each of these requirements below.

\section{Delay--bandwidth limitation}
The achievable time delay $\tau_0$ in a passive linear system (unlike
time-varying active devices~\cite{YanikFa05}) is limited: for a given
bandwidth $\Delta\omega$ and diameter $d$ of the delay region, the
maximum delay is proportional to $d/\Delta\omega$.  The scaling of
delay with bandwidth is known as the delay--bandwidth product
limitation~\cite{Lenz01}, and in the case of a single resonant filter
the upper bound on $\tau_0 \Delta\omega$ is of order unity as a
consequence of the Fourier uncertainty
relation~\cite{Pinsky02,Lenz01}. To obtain a bandwidth much larger
than $1/\tau_0$, one can chain multiple filters into a slow-light
delay line, or even forgo slow light and use propagation through a
long region---in any case, the maximum delay is proportional to the
diameter of the region.  A more careful analysis for slow-light delay
lines yields a delay--bandwidth limit of $\tau_0 \Delta\omega/\omega
\lesssim (n-1) 2d / c$, where $n$ is the effective index in the delay
region~\cite{Tucker05}, and a more optimistic bound of $n (n-1) 2d/c$
was derived under more general assumptions~\cite{Miller07}.  As a
consequence of this and $\tau_0 > 2h/c$, the cloak thickness $d$
must grow proportional to $h$:
\begin{equation}
  d \gtrsim \frac{h}{n(n-1)} \frac{\Delta\omega}{\omega}.
\label{minsize}
\end{equation}
[This is probably optimistic in the wide-bandwidth regime where slow
  light is difficult to utilize; for a non-slow cloak of thickness
  $d$, where the time delay is simply $2dn/c$ and must be
  $> 2(h+d)/c$, one obtains a minimum thickness $d > h/(n-1)$.]
One can relax this tradeoff if a larger
$n$ can be obtained, but large indices of refraction (arising from
resonances) are associated with narrow bandwidths and/or large
losses~\cite{Jackson98}.

\section{Delay--loss limitation}
In the time-delay regime, a larger object for a given bandwidth means
that the incident wave needs to spend more time in the cloak, which
will tend to increase losses due to absorption and imperfections.  The
loss per time $\gamma$ is proportional to $\gamma \sim \omega \Im n /
\Re n$ for light confined mostly in a given
index~$n$~\cite{JoannopoulosJo08-book}.  To maintain effective
invisibility, the loss incurred in the cloak must be small:
one must have $\tau_0 \gamma \ll 1$ for negligible
absorption.  But, since $\tau_0 > 2(h+d)/c$, this implies the
following limitation on the loss tangent:
\begin{equation}
\frac{\Im n}{\Re n} \ll \frac{1}{4\pi} \frac{\lambda}{h+d}.
\label{loss}
\end{equation}
That is, less and less loss can be tolerated
for larger objects relative to the vacuum wavelength~$\lambda$.

In the \emph{phase-delay} regime, the dwell time inside the cloak
can be independent of $h$, in which case the loss tolerance does
\emph{not} decrease as $h/\lambda$ increases, at the expense of
greatly reduced bandwidth.

\section{Interface reflections}
A low-loss cloak achieving the requisite time delay is useless if
there is substantial reflection off the surface of the cloak itself.
In 1d, eliminating reflections 
reduces to the problem of impedance-matching the cloak with
vacuum~\cite{Stratton41}. In the transformation-optics approach,
impedance-matching is attained automatically: the 1d transformation
results in a cloak material that has a both a permittivity
$\varepsilon$ and a permeability $\mu$ (for polarizations transverse
to the surface normal), such that the impedance $\sqrt{\mu/\varepsilon}$
exactly equals that of vacuum~\cite{Ward96}. Alternatively, if the material is
varying slowly enough, this $\mu$ can be approximately commuted with
the curls in Maxwell's equations to combine it with $\varepsilon$ into
an index $n = \sqrt{\varepsilon \mu / \varepsilon_0 \mu_0}$.
This is equivalent to an anti-reflective (AR) coating formed by a
slowly varying $n$ (in the ``adiabatic'' limit of slow variation
the reflection vanishes~\cite{Johnson02:adiabatic}).

This means that a homogeneous medium cannot be used
for the cloak. To obtain a $\mu$,
metamaterials employing subwavelength metallic resonances are
typically used~\cite{Pendry99,Schurig06}, whereas a continuously varying
$n$ is typically achieved with a microporous structure whose
porosity is gradually varied~\cite{KuoPo08, Gabrielli09, Valentine09,
  ErginSt10}. In either case, the loss limit in the previous section
must then include fabrication disorder and surface roughness in
addition to absorption.

\section{Examples and results}
 Let us take some real-world examples of cloaking applications and
 study what practical limitations one would face even for an idealized
 1d ground-plane cloak. For microwave frequencies, consider cloaking a
 vehicle of height $\approx 2\,\mathrm{m}$ from a radar of wavelength
 $\approx 1.25\,\mathrm{cm}$ (24~GHz). Using a time-delay cloak of
 thickness $10$~cm over a $10\%$ bandwidth, \eqreftwo{minsize}{loss}
 imply an effective index of $\gtrsim 1.4$ and a loss tangent of $\ll
 4.7\times 10^{-4}$. (Operating in the phase-delay regime would imply
 a fractional bandwidth of~$< 10^{-4}$.)  [Although one might expect a
   cloak of thickness $h/20$ to need $n=20$ for the requisite delay,
   \eqref{minsize} assumes that slow-light/resonances are used to
   exploit the narrow $\Delta\omega$.] To time-delay cloak the same
 object at visible frequencies with a $10$~cm cloak, aiming for $10\%$
 bandwidth around $575\,\text{nm}$, we would again need $n \gtrsim
 1.4$, but with a loss tangent~$\ll 2 \times 10^{-8}$. (In the
 phase-delay regime, the bandwidth would be only $0.013$~pm.)
 Although such low losses may seem attainable, e.g. with oxides, even
 in 1d a microstructured medium is required for impedance-matching as
 described above, and in 2d and 3d even more complicated metamaterials
 seem
 necessary~\cite{Li08,Liu09,KuoPo08,Gabrielli09,Valentine09,ErginSt10}
 (anisotropy requirements can be minimized via quasiconformal
 transformations~\cite{Li08}, although discarding anisotropy
   incurs a lateral shift in reflected beams~\cite{Baile10}). For a
 cloak at $\approx10$~GHz, an experimental absorption loss tangent
 $\approx 10^{-3}$ was obtained~\cite{Schurig06} for a
 wavelength-scale object; this is already too lossy for 1d cloaking a
 meter-scale object, from above. A ground-plane cloak can use
 non-resonant micro-structures that may be
 lower-loss~\cite{Li08,Liu09}, but for an object that stands many
 wavelengths above the ground, the problem of cloaking against oblique
 waves seems to approach isolated-object cloaking.  Nevertheless, we
 cannot say that the loss bounds from the 1d cloaking problem are
 definitely unattainable for cloaking meter-scale objects at microwave
 frequencies, although it appears challenging.  On the other hand,
 loss tangents $\ll 10^{-8}$ seem impossibly small for any
 metamaterial with metallic constituents at infrared or visible
 frequencies. Even if ground-plane cloaking permits the use of purely
 dielectric constituents, such a loss tangent appears almost
 unattainable when scattering from fabrication disorder and nonzero
 gradients (non-adiabaticity) is included, since the requisite
 gradients (especially for cloaks not too much bigger than the cloaked
 object) seem to imply constituent materials with large index
 contrasts (oxide/air or
 larger)~\cite{Gabrielli09,Valentine09,ErginSt10}.  For comparison, a
 waveguide with a loss tangent of $10^{-9}$ at $1\,\mu$m wavelength
 would correspond to decay lengths of $\sim 1$~km---orders of
 magnitude better than the cm-scale decay lengths typically achieved
 at infrared frequencies in geometries (such as strip waveguides) with
 wavelength-scale geometric components, and a metamaterial requires
 components much smaller than the wavelength (thus many more
 surfaces).  Visible-wavelength cloaking, therefore, seems restricted
 to cloaking objects that are many orders of magnitude smaller than
 meter scales.

\begin{figure}[t]
 \centering
 \includegraphics[width=0.4\textwidth]{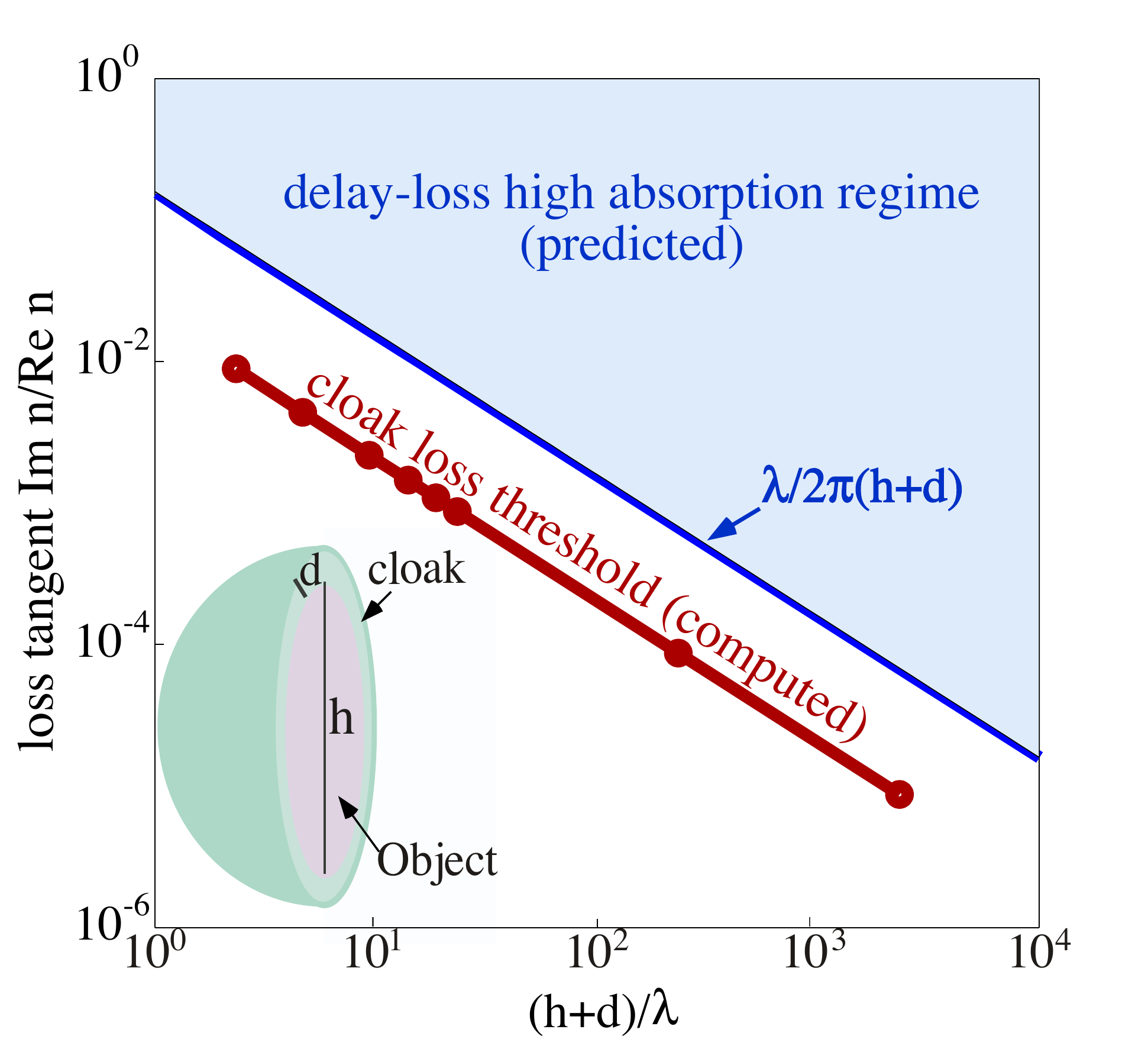}
 \caption{Maximum cloak loss tangent versus diameter $h$ for cloaking
   a perfectly conducting sphere, for cloak of thickness
   $d=h/12$. Shaded area is the regime of high absorption predicted by
   the simple 1d model of \eqref{loss}. The red curve, data from
   \citeasnoun{Baile09}, shows the maximum loss tangent to obtain
   $99\%$ reduction in the scattering cross section using a
   Pendry-type cloak.}
 \label{fig:baile-vs-theory}
\end{figure}

 These calculations demonstrate the difficulty of cloaking objects
 much larger than the wavelength when the ambient medium is
 air/vacuum.  The problem may become easier if the ambient medium
 itself is lossy, such as for an object immersed in water or inside a
 lossy waveguide.  In that case, the loss of the cloak need only be
 comparable to that of the surrounding medium.  The delay--bandwidth
 constraint remains, however: the cloak thickness must grow
 proportional to that of the object being cloaked, for a fixed
 bandwidth. However, if the velocity of light in the ambient medium is
 $< c$, the causality constraint on wide-bandwidth cloaking of
 isolated objects~\cite{Pendry06, Miller06} is lifted. A possible
 example can be found in \citeasnoun{Smolyaninov09PRL}, which cloaked
 an ``object'' (a place where two surfaces touched) roughly 100
 wavelengths in diameter (this ``diameter'' was
 indirectly measured and may not be comparable to the diameters
 of objects used in other cloaking problems), but did so in a
 waveguide between two metallic surfaces.  Such a waveguide has
 a group velocity $< c$, eliminating the causality constraint, and
 may also have non-negligible absorption loss. In addition, the
 cloaking region in \citeasnoun{Smolyaninov09PRL} was achieved by
 curving the surfaces smoothly, which allows a smooth variation of the
 effective index without microstructured media---it seems
 plausible that such a cloak has at least 100 times lower
 absorption/scattering loss than was present in metamaterial
 cloaks.

Although we presented the basic delay--loss and delay--bandwidth/size
tradeoffs in the context of an idealized 1d cloaking problem, we
  believe that similar tradeoffs must apply even more strongly to the
  more difficult problem of cloaking objects in 2d and 3d, especially
  without a ground plane. In fact, recent numerical experiments have
shown that similar tradeoffs (loss tolerance scaling inversely with
diameter) arise for three-dimensional cloaking of a perfectly
conducting sphere of diameter~$h$~\cite{Baile09}. In
\figref{baile-vs-theory}, we show the loss threshold vs. $h$ for this
3d cloak when the (single-$\omega$) reduction in scattering cross
section is fixed to $\nicefrac{1}{100}$, for cloak thickness $d=h/12$.
The scattering is calculated with a transfer-matrix method in a
spherical-wave basis~\cite{Baile09, Qiu09}. Not only does the maximum
loss scale exactly inversely with the diameter, but the
constant factor in this relationship is consistent with the
requirement \eqref{loss} that the loss be much smaller than
$\lambda/2\pi (h+d)$ (shaded region) derived from the much simpler 1d
model (using a path length $h$ instead of $2h$ since there is no
ground plane to double the optical path).

This work was supported in part by the Army Research Office through
the Institute for Soldier Nanotechnologies (ISN) under contract
W911NF-07-D-0004. We are grateful to the referees for helpful comments.


\end{document}